\documentclass[%
 aip,
 amsmath,amssymb,
 reprint,%
]{revtex4-1}

\usepackage{graphicx}% Include figure files
\usepackage{dcolumn}% Align table columns on decimal point
\usepackage{bm}% bold math
\usepackage[utf8]{inputenc}
\usepackage[T1]{fontenc}
\usepackage{mathptmx}
\usepackage{etoolbox}
\usepackage{mathrsfs}
\usepackage{xcolor}

\newcommand{\R}{\textcolor{black}}

\makeatletter
\def\@email#1#2{%
 \endgroup
 \patchcmd{\titleblock@produce}
  {\frontmatter@RRAPformat}
  {\frontmatter@RRAPformat{\produce@RRAP{*#1\href{mailto:#2}{#2}}}\frontmatter@RRAPformat}
  {}{}
}%
\makeatother
\begin{document}

\preprint{AIP/123-QED}

\title[QIS@APS]{\R{Confirming X-ray Parametric Down Conversion by Time-Energy Correlation}}
% Force line breaks with \\
\author{N. J. Hartley}
\email{njh@slac.stanford.edu}
\affiliation{High Energy Density Science Division, SLAC National Accelerator Laboratory, Menlo Park, CA 94025}
\author{D. Hodge}
\affiliation{Department of Physics and Astronomy, Brigham Young University, Provo, UT 84602}

\author{T. Buckway}
\affiliation{Department of Physics and Astronomy, Brigham Young University, Provo, UT 84602}

\author{R. Camacho}
\affiliation{Department of Electrical and Computer Engineering, Brigham Young University, Provo, UT 84602}
\author{P. Chow}
\affiliation{HPCAT, X-ray Science Division, Argonne National Laboratory, Argonne, IL 60439}
\author{E. Christie}
\affiliation{Department of Electrical and Computer Engineering, Brigham Young University, Provo, UT 84602}

\author{A. Gleason}
\affiliation{High Energy Density Science Division, SLAC National Accelerator Laboratory, Menlo Park, CA 94025}
\author{S. Glenzer}
\affiliation{High Energy Density Science Division, SLAC National Accelerator Laboratory, Menlo Park, CA 94025}
\author{A. Halavanau}
\affiliation{Accelerator Research Department, SLAC National Accelerator Laboratory, Menlo Park, CA 94025}
\author{A. M. Hardy}
\affiliation{Department of Physics and Astronomy, Brigham Young University, Provo, UT 84602}
\author{C. Recker}
\affiliation{High Energy Density Science Division, SLAC National Accelerator Laboratory, Menlo Park, CA 94025}

\author{S. Sheehan}
\affiliation{Nevada National Security Sites, North Las Vegas, NV 89031}
\author{S. Shwartz}
\affiliation{Physics Department and Institute of Nanotechnology, Bar-Ilan University, Ramat Gan 52900, Israel}

\author{H. Tarvin}
\affiliation{Nevada National Security Sites, North Las Vegas, NV 89031}

\author{M. Ware}
\affiliation{Department of Physics and Astronomy, Brigham Young University, Provo, UT 84602}

\author{J. Wunschel}
\affiliation{Nevada National Security Sites, North Las Vegas, NV 89031}

\author{Y. Xiao}
\affiliation{HPCAT, X-ray Science Division, Argonne National Laboratory, Argonne, IL 60439}

\author{R.L. Sandberg}
\affiliation{Department of Physics and Astronomy, Brigham Young University, Provo, UT 84602}
\author{G. Walker}
\affiliation{Nevada National Security Sites, North Las Vegas, NV 89031}

\date{\today}

\begin{abstract}
We present measurements of X-ray Parametric Down Conversion at the Advanced Photon Source synchrotron facility. Using an incoming pump beam at 22 keV, we observe the simultaneous, elastic emission of down-converted photon pairs generated in a diamond crystal. The pairs are detected using high count rate silicon drift detectors with low noise. Production by down-conversion is confirmed by measuring time-energy correlations in the detector signal, where photon pairs within an energy window ranging from 10 to 12 keV are only observed at short time differences. By systematically varying the crystal misalignment and detector positions, we obtain results that are consistent with the constant total of the down-converted signal. \R{Our maximum rate of observed pairs was 130 /hour, corresponding to a conversion efficiency for the down-conversion process of $5.3\pm0.5 \times10^{-13}$.}
%While the rates are easily sufficient to be seen above background emission and scattering, we find that they remain significantly lower than theoretical predictions, which we ascribe to \dots
\end{abstract}

\maketitle

\section{Introduction}

Spontaneous Parametric Down Conversion (SPDC) is a nonlinear, quantum optical process \cite{Scully1997,Knight2007,Walls2008}, in which a single (`pump') photon elastically produces a pair of lower-energy photons (generally termed the `signal' and `idler'). 
In the optical regime, where the process was first observed, this process requires a nonlinear (birefringent) crystal such as lithium niobate \cite{Harris1967,Magde1967}. 
The resulting photon pairs have been demonstrated to show properties of quantum entanglement \cite{Kwiat1995}, including violating the Bell Inequalities \cite{Bell1964}. 

SPDC has also been demonstrated experimentally in the X-ray regime. In this case, a traditional nonlinear crystal is replaced by a high-quality low-z crystal, \cite{Freund1969,Levine1970} where a plasma-like nonlinearity can be excited close to a Bragg peak \cite{Shen1984}. This plasma-like nonlinearity is achieved because the incoming x-rays have significantly higher energy than the binding energies of the electrons with which they interact.
%as the momentum and energy conservation can instead be fulfilled by exciting a quadratic nonlinear susceptibility close to a Bragg peak
This was verified in 1971, using scintillator detectors and a Mo-K$\alpha$ source \cite{Eisenberger1971}, which observed down-converted pairs at a rate of around 1 pair/hour. \R{More recent experiments using a similar approach have increased this to around 90 pairs/hour \cite{Shwartz2012}.}

For both X-ray and optical photons, this spontaneous process occurs with very low probability. Unlike other nonlinear processes such as sum frequency generation, which require multiple incoming photons, SPDC is a single-photon-in process.  In the low gain regime, the total count rate of the emitted photons therefore scales linearly with input intensity. Thus, the cross-section of the process cannot be improved by tighter focusing or shorter pulses, with the highest recorded yield rate on the order of $1\times10^{-6}$, seen in a nonlinear niobate crystal with infrared photons \cite{Bock2016}. \R{Nevertheless, the rate of X-ray pairs generated in SPDC exceeds that predicted for other proposed approaches, such as interaction with quantum vacuum fluctuations \cite{Schutzhold2008} or relativistic electrons in an undulating magnetic field \cite{Zhang2023}, and requires a significantly simpler setup than generating pairs by nuclear forward scattering \cite{Palffy2009}.}

While achieving a substantial yield of downconverted X-rays for practical applications is expected to remain challenging, various potential applications for these heralded photons have been proposed. Possible applications include imaging or probing, where the correlations of the photon pairs could allow the background to be significantly reduced \cite{Borodin2016},  or enable imaging with considerably lower doses on sample \cite{Schori2018}. Additionally, previous research has shown how the down-conversion process itself could allow valence charge states to be probed \cite{Boemer2021}.  This could also lead to the development of quantum optical exploration in the X-ray regime, such as by using parametric down-conversion to demonstrate the Hong-Ou-Mandel interference effect \cite{Volkovich2020}, as in the original optical work \cite{Hong1987}. 

In this work, we have demonstrated X-ray SPDC at the \R{Advanced Photon Source (APS)} synchrotron, beamline 16 ID-D. By looking at time-energy correlations of the emitted photons, we are able to confirm that they are produced by down-conversion, and we observe the expected behavior with misalignment angle, which is important to understand for possible applications of the down-converted pairs.

\section{Experiment}

\subsection{X-ray Parametric Down Conversion}

\begin{figure}
\includegraphics[width=0.5\textwidth]{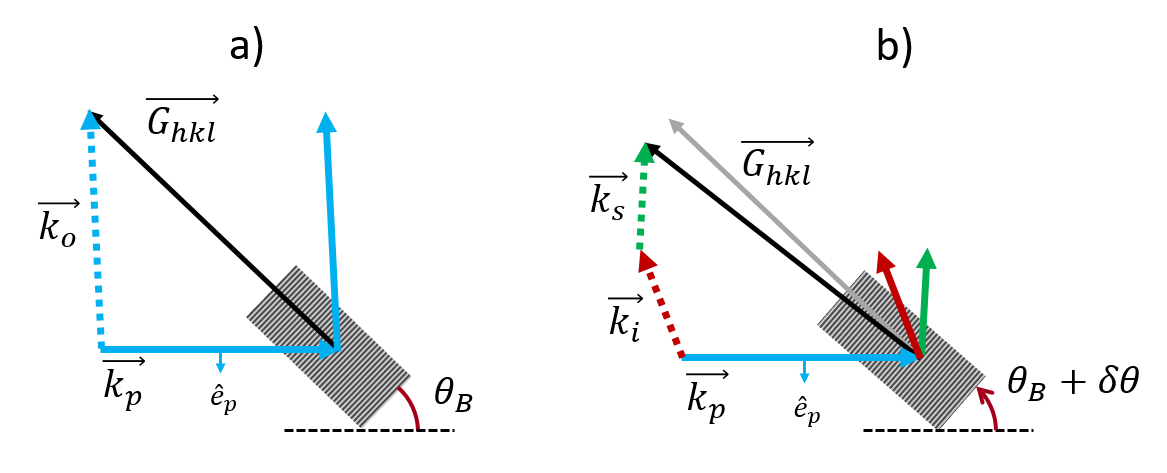}
\caption{\label{fig:SPDC} Schematic illustration of pump-crystal interaction in the case of a) Bragg diffraction and b) parametric down-conversion, around the crystal reciprocal lattice vector with Miller Indices $[hkl]$. In the latter case, signal and idler photons are produced, travelling in directions $\Vec{k_s}$ and $\Vec{k_i}$. \R{The dotted lines indicate the vector calculations, and }$\hat{e}_p$ the polarization of the incoming pump beam.}
\end{figure}

Pairs of photons were produced by down conversion of pump photons with energy $\hbar \omega_p = 22$ keV, in a nominally perfect crystal; in this case, diamond. As shown in Figure \ref{fig:SPDC}, when a crystal in Laue diffraction geometry is detuned by an angle $\delta \theta$  from the Bragg angle, a pair of down-converted photons can be generated, satisfying the phase matching condition:
\begin{equation}
\Vec{k_p} + \Vec{G_{hkl}} = \Vec{k_s} + \Vec{k_i}
\end{equation}
Since the process is elastic, we also require conservation of energy. If we assume that the refractive index for all of the involved photons is 1 (neglecting variations on order of $10^{-7}$ in the x-ray regime), energy and wavenumber are proportional and so this can be expressed as:
\begin{equation}
\left|\Vec{k_p}\right| = 
\left|\Vec{k_s}\right| + \left|\Vec{k_i}\right|
\end{equation}

This process therefore generates two photons with energies $x\hbar\omega_p$, $y\hbar\omega_p$ where $x+y=1$, which we designate the signal, $s$, and idler, $i$. The down-converted photons are emitted at angles $2\theta_B \pm R(x)$, when $x=y=0.5$. As derived in Appendix A, if the misalignment $\delta \theta$, and hence $R(x)$, is small, the deviation from the Bragg angle, R(x), can be well-approximated by \cite{Freund1969,Levine1970,Eisenberger1971}:
\begin{equation}\label{eq:Rx}
R(x) = \sqrt{ 2\delta \theta_B \left(\frac{1-x}{x}\right) \sin \, 2\theta_B }
\end{equation}

The total cross-section of down-conversion is independent of the crystal misalignment \cite{Eisenberger1971}, but for larger values of $\delta \theta$ and hence $R(x)$, the signal is spread over a larger solid angle. Although the illustration in Figure \ref{fig:SPDC} only shows the two-dimensional behavior, in reality the pairs are emitted into cones centred on the Laue peak. In the degenerate case with $x=y=0.5$, these are overlapping and have an opening angle $2\times R(x)$. For a fixed detector area, we would therefore expect a reduction in signal with increasing crystal misalignment, scaling as $1/R(x)$, or equivalently $1/\sqrt{\delta \theta}$.

\subsection{Experiment Details}

The experiment was performed at the 16 ID-D beamline at APS. This facility delivered a beam at 22 keV, with the bandwidth reduced to 2.9 eV FWHM by a Si [111] monochromator. The full beam size was 1.5 mm (horizontal) by 2.5 mm (vertical), and to ensure the beam would not exceed the spatial dimensions of the sample or deviate from the sample during crystal rotations, slits were placed upstream, limiting the beam size to 1 mm by 1 mm. In this measured 1 mm by 1 mm spot, the rate of pump photon arrival was calculated using OASYS (OrAnge SYnchrotron Suite) to be 0.98 × 10$^{13}$ photons per second (2.47 x 10$^{13}$ for full beam size).  The diamond target, with the [110] lattice direction aligned out of the crystal edge, was mounted on a precision rotation stage (SmarAct SR-7012) at Target Chamber Center.  

\begin{figure}
\includegraphics[width=0.45\textwidth]{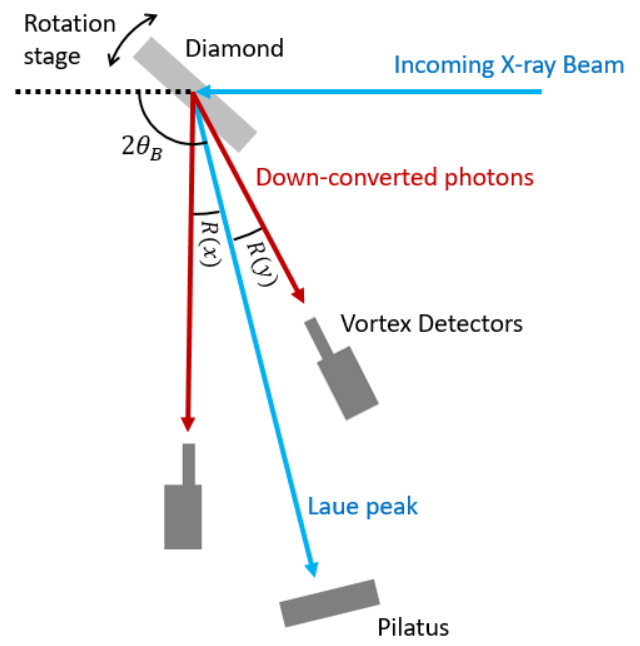}
\caption{\label{fig:layout} Schematic detector layout (from above) showing the area detector (Pilatus, used for alignment) and the energy-resolving Vortex detectors. Note: distances and angles are not to scale.}
\end{figure}

Figure \ref{fig:layout} shows a schematic layout of the detectors used for this experiment. The diffraction spot was detected on an area detector (Dectris Pilatus 100k), and the Hitachi Vortex®-60EX Silicon Drift detectors were positioned to either side of it, at angles calculated by Eq. \eqref{eq:Rx} based on the selected crystal detuning $\delta \theta$. For $2 \theta_B=84.1^\circ$ and $\delta \theta = 10$ mdeg, this gives $R(x) = 1.07^\circ$ in the degenerate $x=y=0.5$ energy split case. The Vortex detectors have 50 mm$^2$ active area and the sample to detector sensor distances were 1351 mm and 1560 mm for the close and far detectors, respectively. We also used a custom helium purge tube with length about 1100 mm with 20 micron thick kapton windows. This tube was made out of 6 inch diameter PVC pipe with an inner diameter of 145 mm. From the two Vortex detectors and the XIA xMAP, the energy ( $\sim$150 eV resolution) and the detection time ( $\sim$20 ns resolution) of photons entering each detector were recorded. 

For simplicity, we measured in the horizontal plane defined by the incoming beam and the diamond (660) reciprocal lattice vector, which was confirmed by scanning the diamond crystal in  $\chi$ (rotation around the incoming X-ray beam direction). Since the beam is horizontally polarized, this reduces the elastic and Compton scattering towards the detectors, by a factor of \cite{Glenzer2009} 

\begin{equation}
    I/I_0 = \left( 1 - \sin^2 2\theta_B \, \cos^2 \chi \right) 
\end{equation}
Thus, the unwanted scattering from these sources is significantly suppressed, consequently reducing detector dead time and giving a greater chance of observing the low-probability down-converted pairs.

\section{Analysis}

\begin{figure*}
\includegraphics[width=0.95\textwidth]{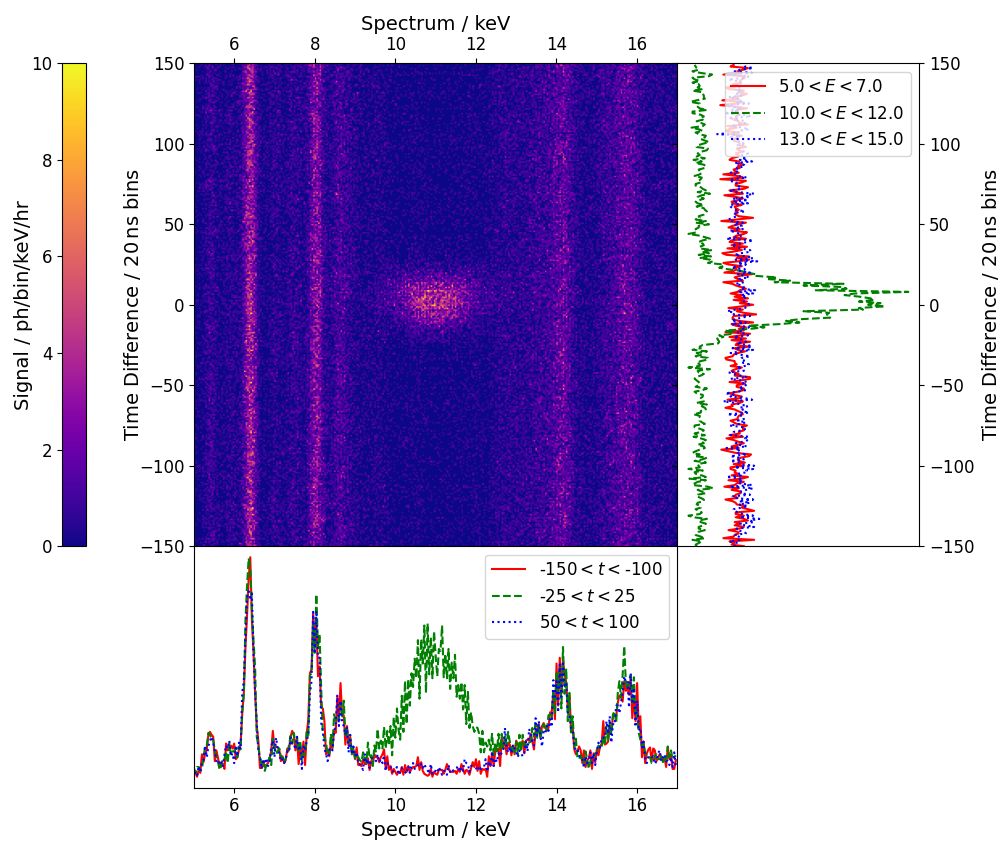}
\caption{\label{fig:ET} Energy resolved Vortex spectra, showing the number of photon pairs (one seen on Vortex 1 at $t1$ with energy $E1$, and another on Vortex 2 at $t2$ with energy $E2$), with $E1+E2$ = 22$\pm$0.5 keV, plotted as a function of ($E1$, $t2-t1$).
%For each time delay between subsequent pairs of photons (y-axis), the image shows the spectrum of detected photons with that time delay. 
This energy-time correlation shows a clear down-converted peak centred at 11 keV, equal to half of the incoming photon energy, which only appears for small time differences. The vertical lines are due to fluorescence and do not change with the time delay between detectors The lineouts show the integrated signal through the central spot (green dashed), as well as the background signal on either side of it (below/left in red solid, above/right in blue dotted).}
\end{figure*}

\subsection{Confirming Parametric Down Conversion}

For each chosen crystal misalignment and corresponding Vortex detector positions, we measured a time sequence of photons and their energies registered in each detector. %In Figure \ref{fig:vspectrum} we show an example of the spectra observed on each detector, averaged over a 9 hour run. 
In order to reduce the signal for analysis, we consider only photons within the energy range of 5-17 keV. Although this suggests that we might miss photon pairs generated with highly asymmetric energy splits, it is unlikely that the photons in those cases would simultaneously strike the detectors at the calculated positions. Additionally, at the lower photon energies, absorption from the air and crystal will reduce the number of idler photons that make it to the detector significantly. 

In order to confirm that we are observing SPDC, we first look at the total energy and time difference for pairs of photons detected on each detector. For the parametric process we are considering, the total energy of the photons must equal that of the pump photon, which is the energy conservation condition from Eq. 2, within the energy resolution of the Vortex detectors i.e. $\hbar \omega_s + \hbar \omega_i = 22\pm 0.5$ keV. Considering only pairs which fulfill this condition, we can then plot the spectrum of these potentially down-converted photons as a function of time difference, which is shown in Figure \ref{fig:ET}. In this figure, the vertical axis is the time delay between measuring a photon on Vortex 1 vs. Vortex 2, while the horizontal axis shows the spectrum of the photons measured with that time difference.

The vertical lines which appear on the figure are due to characteristic x-ray fluorescence emission lines. The signal is the same for any inter-detector time difference, and so is not due to a down-conversion process.  The strongest lines, at 6.4 and 8 keV, are k-$\alpha$ lines of iron and copper, respectively, since these materials are present in beamline components. The higher-energy lines, on the other hand, are primarily due to the requirement that photon pairs of interest sum to $22\pm0.5$ keV; due to the strong lines on the left of the figure, any noise or inelastic signal close to the corresponding points on the right will be exaggerated in the analysis.

The down-converted photon coincidences can clearly be seen in the middle of the figure, centred at 11 keV and 0 ns time difference. The observed spread of time intervals, fitted as a Gaussian distribution with $\sigma = $10.6 bins (212 ns, FWHM = 500 ns), is due to the combined effects of the drift time and peaking time on the two Vortex detectors.

In contrast, the spread in energy is not primarily due to the precision of the Vortex detectors, which at the peaking time chosen is of order 100 eV. Rather, the size of the Vortex active area means that each detector spans a range of in-plane angles and, from Eq. \eqref{eq:Rx}, therefore also a range of energy splits. The width of the peak could therefore be narrowed by using a smaller detector, with a commensurate loss of flux, or by moving to a pixel detector, where the position vs. energy relationship could potentially be observed. However, such detectors would come at the cost of reduced time and energy resolution, and a lower frame rate.
%but the higher capacitance would lead to a lower quantum efficiency.

%\begin{figure}
%\includegraphics[width=0.45\textwidth]{./Figures/Figure_EDiff}
%\caption{\label{fig:energy} Distribution of energy differences between pairs of photons observed with one on each detector. Close to zero time delay, there is a clear peak around 11 keV (half of the incoming pump energy). This is not present at other time delays, and hence does not arise from a fluorescence or other background source.}
%\end{figure}

%Conversely, we can look at the observed energy spectra for pairs of photons within the short delay peak, compared to those for photons arriving at uncorrelated times. Figure \ref{energy} again shows that, as would be expected for down conversion, the photons arriving simultaneously (within the uncertainty of the Vortex detectors) show a peak centered close to 11 keV, corresponding to half the energy of the pump photon. Photon pairs with larger arrival time differences do not contain this same peak, even as other spectral features, corresponding to emission lines, are present.

\subsection{Down Conversion Rates}

\begin{figure}
\includegraphics[width=0.45\textwidth]{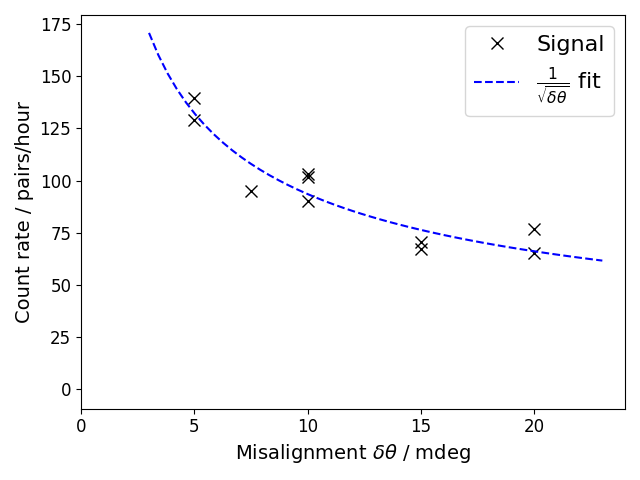}
\caption{\label{fig:analysis} Observed down-conversion rates as a function of misalignment. The observed signal is well-fit by a $1/\sqrt{\delta\theta}$ scaling, which is explained solely by geometric effects and therefore implies a constant down-conversion probability.}
\end{figure}

The rate of down-converted pair generation is found by integrating the signal in the central region of interest of the time-energy graph, as in Figure \ref{fig:ET}, and subtracting the signal rate within the same energy range at other time delays. Next, this is normalized by considering the duration of the run and accounting for fluctuations in the beam current of the APS synchrotron. Although this does not exhibit a perfect correlation with the incoming X-ray flux, the resulting uncertainty attributable to this effect is estimated to be at the percent-level.

Figure \ref{fig:analysis} shows the variation in observed signal rate (pairs/hour) as a function of misalignment. In each case, the pair of Vortex detectors were moved to the expected emission angles, calculated from Eq. \eqref{eq:Rx}, to maximize the observed signal. To confirm the background level, runs were taken with the crystal detuned by -50 mdeg, such that no down-conversion is possible, and the detectors positioned for 10 mdeg detuning. The observed rate of coincidences within the same region of interest as above was <1 count/hour, compared to ~100/hour with positive detuning. \R{Given the incoming photon rate, this suggests that we observe one pair for every $2.7 \times 10^{14}$ photons incident on the crystal.}

We see that the rate of observed pairs falls with increasing crystal misalignment. As explained above, this is due to the geometry of our setup, where a fixed detector area covers a smaller proportion of the down-scattering ring at larger misalignment. Our results are therefore consistent with a down-conversion rate which does not change with crystal misalignment but with this geometrical effect of looking at different portions of the SPDC cone. \R{Accounting for this variable coverage, the rate of observable pair generation is $3400 \pm 300$ pairs/hour for each misalignment studied.}

Theoretical estimates \cite{Freund1969, Levine1970} suggests that pairs should be \R{incident on our detectors} at a rate of around 70 pairs/minute for a detuning of 10 mdeg. \R{This is reduced by around 82\% due to} the combined effects of air absorption between the crystal and detector, and detector inefficiency due to the high penetrating power of the X-rays. However, our observed rates are nearly a further order of magnitude below this. Finding rates so far below theoretical estimates is not unusual for parametric down-conversion experiments \cite{Borodin2016}, and our observed rate is in line with previous results \cite{Shwartz2012,Sofer2019}. Given the observed peak width of the crystal as discussed in Appendix B, we propose that this may be caused by a reduction in the effective crystal thickness due to the mosaicity in the crystal, but further experiments would be required to confirm this.

%We note that this is hard to reconcile with the initial calculations of down-conversion rates by Freund and Levine \cite{Freund1969, Levine1970}. Those papers calculate a rate which scales proportionally to the solid angle $\Delta \Omega$ and inversely with $\delta \theta$. However, 

%As expected from Eq. \eqref{eq:prob}, the rate of down-converted pairs decreases at larger misalignment angles. In contrast to expectations, however, the count rate scales as $(\delta\theta)^{-1/2}$, rather than the expected $(\delta\theta)^{-1}$. Since the pairs were observed in the expected locations, confirming that our calculation of $R(x)$ is correct, this appears to suggest that the probability scales as $P(x)\propto R(x)^{-1}$, as opposed to the $R(x)^{-2}$ relation given in Eq. \eqref{eq:prob} above.

\section{Conclusion}

In conclusion, we have observed the production of X-ray pairs by spontaneous parametric down conversion, in a near-Bragg diffraction setup. That the photons were produced through this process was confirmed using time-energy correlations in the observed photons, which exhibit a clear peak at zero time difference between the detectors. The decrease in observed pair production as the crystal misalignment increases can be attributed to geometric effects, \R{with the observed signal further reduced by losses in the air and inefficiency in the detector. Accounting for these effects, our results indicate a total pair generation rate of $18900 \pm 1700$ pairs/hour, which remains unchanged with misalignment. This corresponds to a conversion efficiency for the down-conversion process of $5.3\pm0.5 \times10^{-13}$}. Our results suggests that future experiments using reflective focusing optics and different detectors could greatly increase the rate of usable down-converted photons.

\begin{acknowledgments}
This work was supported by U.S. Department of Energy (DOE) Office of Fusion Energy Sciences funding Nos. FWP100182 and FWP 100866,  by Mission Support and Test Services, LLC, under Contract No. DE-NA0003624 with the U.S. Department of Energy, the Office of Defense Programs, and supported by the Site-Directed Research and Development Program. DOE/NV/03624--1787, and by funding from the College of Physical and Mathematical Sciences at Brigham Young University. A.H. would like to thank A. Benediktovitch, C. Boemer, and D. Krebs (CFEL, DESY) for insightful discussions on the subject.

The experiment was performed at HPCAT (Sector 16), Advanced Photon Source (APS), Argonne National Laboratory. HPCAT operations are supported by DOE-NNSA’s Office of Experimental Sciences. The Advanced Photon Source is a U.S. Department of Energy (DOE) Office of Science User Facility operated for the DOE Office of Science by Argonne National Laboratory under Contract No. DE-AC02-06CH11357. We gratefully acknowledge support received from the APS Detector Pool, especially Christopher Piatak and Antonino Miceli.
\end{acknowledgments}

\section*{Data Availability Statement}
The data that support the findings of this study are available from the corresponding author upon reasonable request.

\appendix
\section{Emission Angles of Down Converted Photons \cite{Adams2003}}

\begin{figure}
    \centering
    \includegraphics[width=0.45\textwidth]{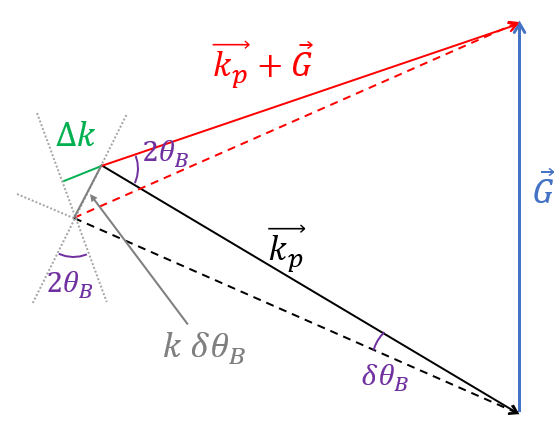}
    \caption{Setup with $\delta\theta_B$ detuning (not to scale). The black and red dashed lines show $\Vec{k_p}$ and  $\Vec{k_p}+\Vec{G}$ before detuning.}
    \label{fig:setup}
\end{figure}

In the initial Bragg diffraction setup, $\Vec{k_p}$ and $\Vec{k_p}+\Vec{G}$ are equal in magnitude, since it is an elastic process. We can therefore define $\Delta k$ to be the change in magnitude of $\Vec{k_p}+\Vec{G}$ when the small detuning angle $\delta \theta_B$ is introduced. As shown in Figure \ref{fig:setup}, this quantity is equal to:
\begin{align}   
    |\Vec{k_p}+\Vec{G}| &= k - \Delta k     \label{eq:delk2} \\
    \Delta k &= k\delta\theta_B\sin{2\theta_B}     \label{eq:delk1}    
\end{align}
where $k=|\Vec{k_p}|$.

\begin{figure}
    \centering
    \includegraphics[width=0.3\textwidth]{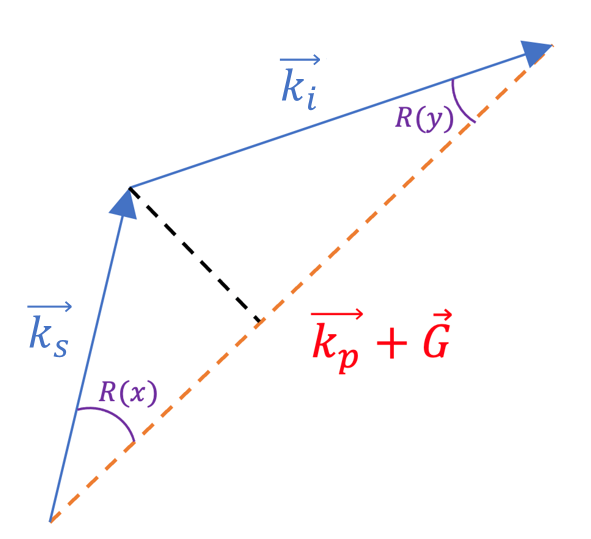}
    \caption{Momentum Conservation of Down Converted Photons}
    \label{fig:pair}
\end{figure}

From conservation of momentum, shown in Figure \ref{fig:pair}, we can define the following relations for the angles of the outgoing photons.
\begin{equation}
    \label{eq:mom1}  |\Vec{k_s}|\sin{R(x)} = |\Vec{k_i}|\sin{R(y)} 
 \end{equation} 
 \begin{equation}
    \label{eq:mom2} |\Vec{k_s}|\cos{R(x)} + |\Vec{k_i}|\cos{R(y)} = |\Vec{k_p}+\Vec{G}| 
\end{equation}

Next, we substitute $|\Vec{k_s}|=xk$ and $|\Vec{k_i}|=yk$ into \ref{eq:mom2} according to the normalized conservation of energy relationship, $x+y=1$.
\begin{equation}
    |\Vec{k_p}+\Vec{G}| = xk\cos{R(x)} + (1-x)k\cos{R(y)}
    \label{eq:intmag1}
\end{equation}

In order to approximate each of these cosine terms, we start with the identity $\cos{R(y)} = \left(1-\sin^2{R(y)}\right)^{1/2}$ and employ our first momentum condition \ref{eq:mom1}.
\begin{align*}
   \cos{R(y)} &= \left(1-\frac{|\Vec{k_s}|^2}{|\Vec{k_i}|^2}\sin^2{R(x)}\right)^{1/2} \\
    &= \left(1-\frac{x^2}{(1-x)^2}\sin^2{R(x)}\right)^{1/2}
\end{align*}
after substituting the normalized magnitudes.

We can then use small angle approximations and a first order binomial approximation in order to obtain:
\begin{align*}
    \cos{R(y)} &\approx 1 - \frac{1}{2}\frac{x^2}{(1-x)^2}(R(x))^2 \\
    \cos{R(x)} &\approx 1-\frac{1}{2}(R(x))^2
\end{align*}

We then substitute these approximations into \ref{eq:intmag1}
\begin{equation*}
\begin{aligned}
    |\Vec{k_p}+\Vec{G}| \approx \,k\Bigg[ x&\left(1-\frac{1}{2}(R(x))^2\right)+ \\
    &(1-x) \left(1 -\frac{1}{2}\frac{x^2}{(1-x)^2}(R(x))^2 \right) \Bigg] 
\end{aligned}
\end{equation*}

Finally, we substitute \ref{eq:delk2} and solve for $\Delta k$,
\begin{equation}
    \Delta k = \frac{1}{2}k(R(x))^2\left(x + \frac{x^2}{(1-x)}\right)
\end{equation}
employ relation \ref{eq:delk1}
\begin{equation*}
    \delta\theta_B\sin{2\theta_b} = \frac{1}{2}\frac{x}{y} \big[ R(x) \big]^2
\end{equation*}
and solve for $R(x)$.
\begin{equation}
    R(x)=\sqrt{2 \delta\theta_B \frac{y}{x}\sin{2\theta_B}}
\end{equation}

\section{Crystal Characterization}

\begin{figure}
\includegraphics[width=0.45\textwidth]{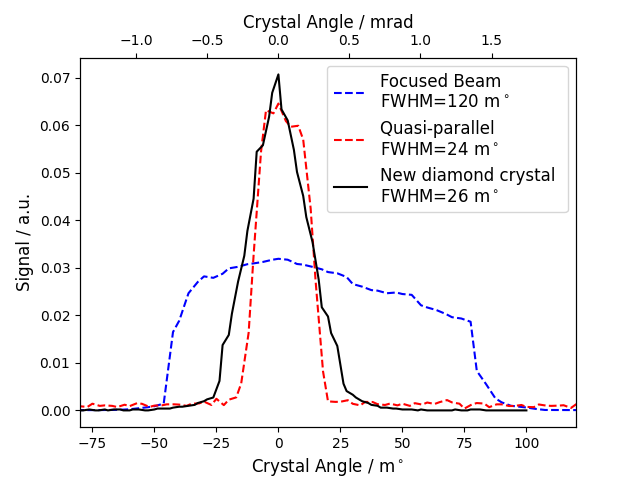}
\caption{\label{fig:anglescans} Peak widths observed as the crystal angle is scanned. The decrease in width as focusing optics are removed is clear.}
\end{figure}

\begin{figure}
\includegraphics[width=0.85\linewidth]{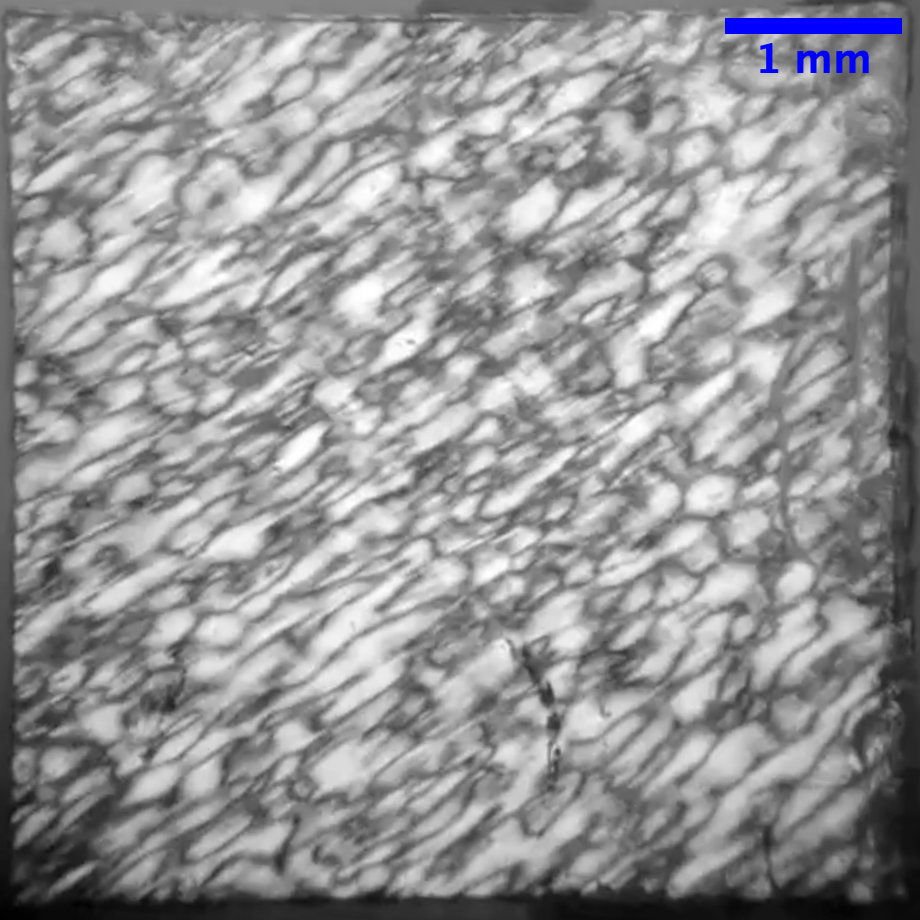}
\caption{\label{fig:diamondcrystal} Image of our diamond crystal taken using a dual-polarizer microscope configuration, with one polarizer fixed while the other is rotated. This confirms that our sample consists of many domains, which would possibly contribute to the peak widths seen during the angle scans.}
\end{figure}

The initial width of the diffraction peak was found by scanning the crystal rotation and observing the change in signal intensity on the Pilatus detector, or on an Si PIN X-ray diode. The width of this peak is due to the combined effects of the focusing divergence of the incoming X-ray beam, the Darwin width of the diffraction peak, and mosaicity in the crystal. Scans were performed in different beam modes, and with two different diamond crystals, and the outputs of the scans are shown in Figure \ref{fig:anglescans}.

The dashed lineouts show data taken with the initial diamond crystal, with the beam initially in a focused mode, and then with the focusing Kirkpatrick-Baez (KB) mirrors flattened as much as possible. This shows a significant decrease in the width of the diffraction peak, although it is still much wider than the expected Darwin width, estimated at $\sim 1 \mu$rad FWHM. The final (black solid) lineout was taken with a different diamond crystal, and with the focusing mirrors removed entirely; in this case, the width of the peak appears to be due to mosaic behavior in the crystal, possibly induced or worsened by strain.

\section{Raw Vortex Data}

\begin{figure}
\includegraphics[width=0.45\textwidth]{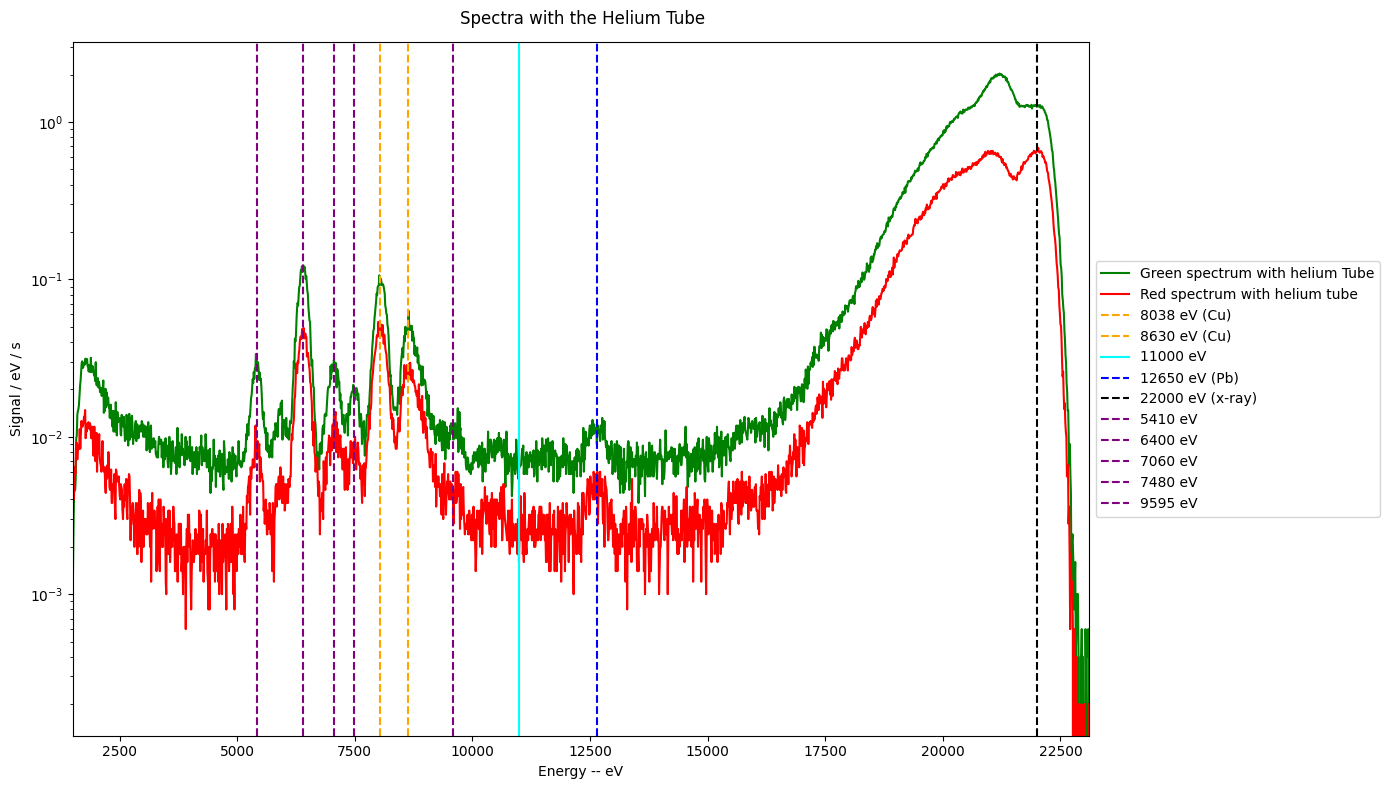}
\caption{\label{fig:Vspec} Raw spectra observed on the Vortex detectors. Even with the detector positions located in the plane of polarization, the signal is dominated by the high energy elastic (diffuse) and Compton peaks. Significant fluorescence peaks are labelled.}
\end{figure}

Since this is a low cross-section process, the raw Vortex data is dominated by other sources of X-ray signal. As can be seen in Figure \ref{fig:Vspec}, this includes thermal diffuse scattering, which leads to the elastic peak at 22 keV, and Compton scattering, leading to the broad down-shifted peak. Lower energy peaks are due to fluorescence lines, labelled in the Figure, primarily from beamline components or shielding around the interaction point and detectors.

\section*{Bibliography}
\nocite{*}
\bibliography{APS_manuscript}% Produces the bibliography via BibTeX.

\end{document}